
%
%
%
%
%
\magnification=\magstep1\overfullrule=0pt\pageno=0\hsize=13.4cm\noindent
\baselineskip=12pt\lineskiplimit=12pt\lineskip=12pt
\nopagenumbers\line{\hfil OCIP/C-92-1}
\vskip1cm
\centerline{\bf Production of the Lightest Supersymmetric Particle}
\vskip.1cm
\centerline{\bf in Electron-Photon Collisions}
\vskip.5cm
\centerline{Heinz K\"onig$^{1*}$ and K. A. Peterson$^2$}
\centerline{Ottawa-Carleton Institute}
\centerline{for Physics}
\centerline{Department of Physics}
\centerline{Carleton University}
\centerline{Ottawa, Ontario}
\centerline{Canada K1S 5B6}
\vskip.5cm
{\bf Abstract}: We present cross sections for the production of the
lightest supersymmetric particle as a neutralino state in the minimal
supersymmetric standard model at electron-photon colliders.
The lightest supersymmetric particle mass is taken at a value of
30 GeV which is slightly higher than its
lowest experimental bound of 20 GeV, and the masses of the scalar electron
are varied. We show partial cross sections of the energy and
angular distribution of the outgoing electron for different values
of the centre of mass energy.
As a result we show that electron-photon collider experiments
could be quite sensitive to the detection of supersymmetric particles.
\hfill\break \vskip.1cm
\centerline{April 1992}
\vskip.0cm\noindent
$\overline{^*\rm supp}$orted by Deutsche Forschungsgemeinschaft\hfill
\sevenrm\break$^1$EMAIL: KONIG@PHYSICS.CARLETON.CA\hfill
\break$^2$EMAIL: PETERSON@PHYSICS.CARLETON.CA\hfill\break\tenrm\eject
\footline={\hss\tenrm\folio\hss}

\centerline{\bf I. INTRODUCTION}\hfill\break\vskip.1cm
In very recent papers [1-7 and references therein] it was shown that
future electron-photon and photon-photon colliders could be quite sensitive to
the observation of new physics beyond the standard model. Experimentally,
electron-photon and photon-photon colliders could be obtained by
backscattering laser beams off $e^+e^-$-colliders such as SLC or NLC (Next
Linear Collider) [8].  One could also consider scenarios where the photon
originates either from bremstrahlung or beamstrahlung [9]. In the literature it
has been shown that electron-photon and  photon-photon colliders could be used
to
limit anomalous gauge couplings [3,4,5], to produce the SM Higgs boson [2], to
produce isosinglet neutral heavy leptons [1] or to produce supersymmetric
particles of the  minimal supersymmetric standard model (MSSM) such as
charginos
and scalar leptons [6,7]. In these papers the centre of mass energy
(CME), $\sqrt{s}$, ranged from 100 GeV to 2 TeV.\hfill\break
\indent In this
paper we present the cross sections for producing the lightest supersymmetric
particle (LSP) and the scalar electron with its following decay into an
electron and the LSP. We suppose that the LSP is the lightest mass eigenstate
of the 4x4 neutralino mass matrix and is mainly a photino. We consider
beam centre of mass energies from SLC like energies of 100 GeV to
possible future energies of 1 TeV. \hfill\break
\indent We present our results in the following
sections. In the next section we lay the ground work for the energy
distribution
of the photon beam that would result from an $e\gamma$\ collider produced by
laser backscattering.  These distributions will then be folded into our
cross-section evaluations. In the third section we show the Feynman diagrams
for
producing the LSP and give the couplings. We also discuss the masses of the
considered supersymmetric particles and take values  consistent with the
experimental lower limits. In section four we present the partial cross
sections
of the angular and energy distributions of the outgoing electron for various
values of the scalar electron mass and CME. In the last section we give our
conclusions. The results of our calculation of the matrix element squared are
given in the appendix. \hfill\break

\vskip.2cm\noindent
\centerline{\bf II. BACKSCATTERING LASERS OFF ELECTRON BEAMS}\vskip.2cm
\indent
The idea of Compton scattering laser light off $e^-$ or $e^+$ beams was
first proposed by Ginzburg et al. [8].  This possibility is especially
exciting, since the scattered  photon beam is hard, with energies and
luminosities   very close to those of the original electron/positron
beam.\hfill\break
\indent The relation between the subprocess $e^-\gamma\rightarrow X$\ and the
$e^+e^-$\ collisions may be approximated by folding its $e\gamma$\ cross
section with the appropriate differential $e\gamma$\ luminosity function[4],
$${{d\sigma_{e^+e^-\rightarrow e^+e^- X}}\over{dcos\theta}} \approx
\int\limits^1_{\tau_{\rm min}} d\tau {{dL_{e\gamma}(\tau)}\over{d\tau}}
{{d\sigma_{e\gamma\rightarrow X}}\over{dcos\theta}},\eqno(1)$$
where $\tau_{\rm
min}=Max(\hbox{lower kinematic limit},x_{\rm max}/2)$\ (see below) and the
differential luminosity $dL_{e\gamma}/d\tau$\ is defined by the momentum
distributions $f_{e,\gamma}$\ of the electron and photon beams respectively,
$${{dL_{e\gamma}}\over{d\tau}} =  \int\limits^{x_{max}}_{\tau /x_{max}}
{{dx}\over{x}} f_e(x)f_{\gamma}(\tau /x),\qquad \tau =
{{\hat{s}}\over{s}},\eqno(2)$$ with s being the total $e^+e^-$ CME squared,
$\hat{s}$\ being the effective $e\gamma$\ CME, and $x_{max}$ corresponding to
the maximum momentum allowed kinematically by the $\gamma$\ production process.
Since the electron beam distribution may be
represented by a $\delta$-function, the
subprocesses are  related to the $e^+e^-$ processes simply by the momentum
distribution for Compton scattering, $${{d\sigma_{e^+e^-\rightarrow e^+e^-
X}}\over{dcos\theta}} \approx \int\limits_{x_{min}}^{x_{max}} d\tau
f_{\gamma}^{laser}(\tau)  {{d\sigma_{e\gamma\rightarrow
X}}\over{dcos\theta}},\eqno(3)$$ where $x_{min}$\ is the minimum allowable
momentum in the phase space of the process under consideration.  The momentum
distribution of the Compton scattered laser photons is just given by the
differential cross-section for Compton scattering[4,8],
$$f_{\gamma}^{laser}(x,x_0) = {1\over{\sigma_C}}{{d\sigma_C}\over{dx}}
= {{1-x+{1\over{(1-x)}}-{{4x}\over{x_0(1-x)}} + {{4x^2}\over{x^2_0(1-x)^2}}}
\over{(1-{4\over{x_0}}-{8\over{x^2_0}})\ln (1+x_0) + {1\over 2} + {8\over{x_0}}
- {1\over{2(1+x_0)^2}} }},\eqno(4)$$
with $x_0$\ being a Compton scattering variable given by
$$x_0 = {{2 \sqrt{s}\omega_{laser}}\over{m_e^2}},\eqno(5)$$
dependent upon the electron beam energy, $\sqrt{s}/2$, the laser energy,
$\omega_{laser}$, and the electron mass, $m_e$.  The maximum photon energy
allowed by the Compton scattering is
$$x_{max} = {{x_0}\over{1+x_0}}.\eqno(6)$$
We will use $\omega_{laser} =$\ 1.17 keV or 3.5 keV,
corresponding to a neodymium glass laser and a neodymium glass
laser with frequency tripling respectively[8].  The back-scattered portion of
the Compton scattering is equivalent to the large x portion.  One can thus
concentrate on the hard scattering of the laser beam by collimating the
back-scattered laser in a small angle.  We have therefore required $\tau >
x_{max}/2$\ when folding in the photon distributions [8].
 \hfill\break\indent

\vskip.2cm\noindent
\centerline{\bf III.
FEYNMAN DIAGRAMS, COUPLINGS AND MASSES}\vskip.2cm To produce the LSP in
electron-photon collisions we have to consider the three Feynman diagrams given
in Figure 1 with the Feynman rules [10] of Figure 2. In the calculation of the
matrix element we have also to  consider the diagrams where $\tilde\chi_i^0$\
is interchanged with $\tilde\chi_j^0$\ in Figure 1, that is $p_1\leftrightarrow
p_2$. The complete matrix element squared is then $\displaystyle{{1\over
2}\vert
M(p_1,p_2)-M(p_2,p_1)\vert^2}$. A detailed calculation of the matrix
element squared is given in the appendix.\hfill\break
\indent The highest background to
this process in the standard model comes from $e\gamma\rightarrow
e\nu\bar{\nu}$, which leads to the same signature of
$e\gamma\rightarrow\ e\ +$\ missing energy.
This process may proceed through a
W or Z propagator [3-5]. \hfill\break \indent For
the masses of the supersymmetric particles we have to take values consistent
with the experimental lower limits. From LEP and SLC experiments it is well
known that the masses of new charged particles such as charginos or scalar
leptons must be larger than 45 GeV. If the LSP is a neutralino, it has been
shown that its mass must be larger than about 20 GeV [11]. We do not consider
the scalar electron neutrino as the LSP.  In the MSSM, due to the D-term in the
superpotential and the renormalization group equations, the relationship
between
the SU(2) doublet scalar leptons is [12]\hfill\break
$$m^2_{\tilde\ell}-m^2_{\tilde\nu_\ell}=m^2_W{{v_2^2-v_1^2}
\over{v^2_2+v^2_1}}.\eqno(7)$$\hfill\break
 With $v_1\approx v_2$\ there is no
difference in the masses of the scalar electron and scalar electron  neutrino.
We also do not consider a large mixing between the scalar partners of the
left- and right-handed electrons, which is proportional to the electron mass in
the MSSM [10].\hfill\break
\indent The 2x2 chargino mass matrix and the 4x4
neutralino mass matrix are described by the higgsino mixing parameter $\mu$,
the
$SU(2)_L$\ gaugino mass $m_{g_2}$\ and the ratio of the Higgs expectation
values $\tan\beta= {{v_2}\over{v_1}}$\ [10]. We use the relation of the
$SU(2)_L$\ and $U(1)_Y$\ gaugino masses $m_{g_2}/m_{g_1}=
{{3\alpha_2}\over{5\alpha_1}}$\ in grand-unified supersymmetric models.
\hfill\break
\indent Analyses of the chargino and neutralino mass matrices
show that negative values of the $\mu$-parameter are favoured. The parameter
region $0\le\mu\le 50$\ GeV leads to too light chargino masses. If $\mu$\ is
very large ($\ge 100$ GeV) then the gaugino mass has also to be very large
(larger than $\mu$) to avoid too small chargino and neutralino masses. If
$\mu$\ is negative, we have to keep  $m_{g_2}\le\vert\mu\vert$ if we want to
have the LSP be mainly a photino (that is $N'_{11}$\  in Figure 2 b)-e) is of
${\cal O}(1)$). In the case $v_1=v_2$\ and  $m_{g_2}\ge\vert\mu\vert$\ the LSP
is
mainly a higgsino with mass $\vert\mu\vert$\ (this can be easily seen by
calculating the polynom for the mass eigenvalues $\lambda$\ and setting
$\lambda=-\mu$\ which  is a pure mass eigenstate if $v_1=v_2$) (here $N'_{11}$\
and $N'_{12}$\ are zero). In the case $v_1\neq v_2$\ the LSP is a
mixture of all components with increasing mass as $m_{g_2}$\ grows.
\hfill\break\indent
In the cross sections given below we have taken a value of 30 GeV for the LSP
mass, a value just slightly above the lowest experimental limit of 20 GeV.
To have not too many parameters we
have also taken it to be primarily a photino ($N'_{11}\approx 1$).  This is
valid for the region $\mu\le 0$\ and $m_{g_2}\le\vert\mu\vert$. For purposes
of calculating the width of the scalar electron, we have also assumed that a
zino
state ($N'_{22} \approx 1$) and a wino state occur at 100 GeV.  Other
neutralino
and chargino states are assumed to be either too heavy or have too weak a
coupling to contribute to the scalar electron width.  In addition to a 30 GeV
photino, we will briefly comment on higher LSP masses. In the next section we
present the cross sections for CME values of $\sqrt{s}=100$\ GeV, 500  GeV and
1
TeV, as well as for different masses of the scalar electron. \hfill\break
\eject\indent

\noindent
\centerline{\bf IV. CROSS SECTIONS FOR THE PRODUCTION OF THE LSP}\vskip.2cm
We present the partial cross sections $\displaystyle{{{d\sigma}\over{dE_e}}}$\
and  $\displaystyle{{{d\sigma}\over{d\cos\theta_e}}}$, where $E_e$\ is the
energy
of the outgoing electron and $\theta_e$\ its angle with the incoming electron.
A beamline cut of 10$^o$ has been made on the outgoing
electron.\hfill\break
\indent In Figure 3 a) we show the energy distribution of the
outgoing electron with a CME for $e\gamma$ collisions of 100 GeV. This is done
for  scalar electron masses of 50, 75 and 100 GeV. The solid line gives the
Standard Model result for the process  $\gamma e\rightarrow e\nu\bar{\nu}$,
while the
dashed line gives the MSSM result for a scalar electron mass of 50 GeV, the
dotted line for $m_{\tilde e}=75$\ GeV and the dash-dotted line for $m_{\tilde
e}=100$\ GeV. We have also included the cross-sections for the case when the
incoming electron has been right-hand polarized. In Figure 3,
only the Standard Model result
differs from the unpolarized case and it is given by the dash-triple-dotted
line.  Figure 3 b) gives the angular distribution of the outgoing electron.
While in the Standard Model the outgoing electrons are predominantly
back-scattered with respect to the incoming electrons, the MSSM has a more even
angular distribution. For a scalar electron mass of about 50 GeV a experimental
cut at $\cos\theta=-0.8$\ could distinguish the MSSM result from the standard
result. As a result we see that scalar electron masses small enough (about  50
GeV) lead to cross sections comparable with the standard result.  If one has a
polarized electron beam, one could  produce cross-sections considerably above
the Standard Model result, there being only a contribution from the Z in the
Standard Model, and this occurs at a very well defined electron energy. Higher
scalar electron masses above the threshold for real scalar electron production
give cross sections that are too small.
 \hfill\break\indent
In Figure 3 c) and d) we consider the same cases as in
 Figure 3 a) and b) with the photon
energy distribution folded in for an $e^+e^-$\ CME $\sqrt s=100$\ GeV
and a laser beam energy of 3.5 keV according to (1). This leads to a further
suppression of both the supersymmetric process and the S.M. background by a
factor of about 10. The shape of the figures is slightly
changed due to a smearing of the effective CME for $e\gamma$\ collisions.
\hfill\break\indent  Figure 4 a) and b)
shows the results for an $e\gamma$\ CME
of 500 GeV and  scalar electron masses of 50, 100 and 250 GeV.  The S.M. result
is given by the solid line while the 50, 100 and 250 GeV scalar electron
differential cross-sections are given by, respectively, the dashed, dotted and
long-dash-dotted lines.  The right-hand polarized results are given for the
S.M.
by the dash-triple-dotted line, and for the 250 GeV scalar electron by the
short-dash-dotted line.  The small difference in the polarized cross-section
for
the 250 GeV scalar electron is due to the difference in decay modes (and thus
decay widths) of the right- and left-handed scalar electrons, the left-handed
scalar electron having a decay channel through the 100 GeV
wino.\hfill\break\indent In Figure 4 c) and d)
we have again folded in the photon energy
distribution with a more realistic laser beam energy of 1.17 keV. Because of
(5) we have taken a somewhat higher laser beam energy for the $e^+e^-$\ CME of
100 GeV. Because of the higher CME  we only get a
small suppression factor in contrast to Figure 3 c) and d). In Figure 4 d)
we see that the MSSM is comparable to the Standard Model
and even can be distinguished if the incoming electron beam is polarized.
\hfill\break\indent  In Figure 5 a) and b)
we have an $e\gamma$\ CME of 1 TeV
and scalar electron masses of 50, 250 and 500 GeV. The same labeling scheme as
the 500 GeV case was used. In Figure 5 c) and d)
the photon energy distribution with the
same laser beam energy as in Figure 4 c) and d)
was folded in. The results are similar to the results shown in
Figure 4 c) and d). \hfill\break\indent
  We observe that
electron-photon colliders at high energies could give strong lower bounds on
scalar electron masses or may even lead to their detection.\hfill\break
\indent Higher values of the LSP mass lead to a further phase space suppression
of the cross section. For an LSP of about 90 GeV the results are not changed
significantly.  Total cross sections change by  no more than a factor of 2,
while the only significant change is in the energy distribution of the outgoing
electrons.  The high cross section portions of these distributions are those
portions for which the electron and the second LSP proceed through a real
scalar electron.  The maximum electron energy for this to happen is given by
$E_e = {1\over 2}(m_{\tilde{e}}^2 - \tilde{m}^2) /
(E_{\tilde{e}}-|\vec{p}_{\tilde{e}}|)$, which decreases for increasing LSP
mass,
$\tilde{m}$. The total cross sections lie in the 10 fb to 1 pb region.
\hfill\break

\vskip.1cm\noindent
\centerline{\bf V. CONCLUSIONS }\vskip.1cm
In this paper we have presented the partial cross sections for the energy and
angular distributions of the outgoing electrons coming from the decay of the
scalar electrons in electron-photon collisions. The missing energy is carried
away by two LSP's. We have given the result for three different values of the
CME, $\sqrt{s}=100$\ GeV, 500 GeV and 1 TeV. The LSP was given a mass of 30
GeV,
just above  the lowest experimental limit of 20 GeV valid for
the LSP being a neutralino. In our analysis we have plotted the differential
cross sections $\displaystyle{{{d\sigma}\over{dE_e}}}$\ and
$\displaystyle{{{d\sigma}\over{d\cos\theta_e}}}$\  for various scalar electron
masses for the case of a $\sqrt{s}/2$\ photon beam incident on a $\sqrt{s}/2$\
electron beam, and for the case where the photon energy distribution
of a backscattered laser has been folded in. \hfill\break\indent As a
result we see that $e\gamma$-colliders could be more  efficient at
constraining the scalar electron and LSP masses (or possibly even detecting
them) than $e^+e^-$-colliders, where the scalar electrons are produced in
pairs.
We have also shown that if the scalar electron mass is small enough, the cross
sections can be distinguished from the Standard Model backgrounds. Folding in
the photon energy distribution with a laser beam energy of 1.17 keV and for an
$e^+e^-$\ CME of $\sqrt s= 500$\ GeV and 1 TeV  we have seen in Figure 4 d) and
5 d) that even for higher values of the scalar electron masses, the MSSM can be
distinguished from the Standard Model if the incoming electron beam is
polarized. Electron-photon colliders could provide us
 with either high constraints on the
mass of the scalar electron and LSP or evidence for supersymmetry.\hfill\break
\indent While finishing our
calculations we received a paper by Goto and Kon [13]. They consider the
production of scalar electrons and photinos and the production of  scalar
electron-neutrinos and charginos. They present the angular distribution of the
outgoing scalar electron, chargino and W-boson. In our paper we have
considered the full cross section for producing
two LSP's and an electron through an intermediate LSP/scalar electron state.
Thus we present the angular and energy distribution of the outgoing
electron coming from the decay of the scalar electrons.  It is these quantities
that would be experimentally observable.  We believe this is necessary so that
one can properly distinguish this event from the SM background.  In
the limit that our results should agree with those of [13], they
do.\hfill\break  \vskip.1cm\noindent \centerline{\bf IV.
ACKNOWLEGEMENTS}\vskip.1cm The authors would like to thank S. Godfrey and R.
Sinha for help with the Monte Carlo phase space integrations.  The work of H.K.
was supported by the Deutsche Forchungsgemeinschaft, while that of K.A.P. was
supported through an NSERC grant. \hfill\break \vskip.1cm\noindent
\centerline{\bf IV. APPENDIX}\vskip.1cm In this appendix we give the result of
the calculation of the matrix elements for the Feynman diagrams in Figure 1. We
have done the calculation for the matrix elements in independent ways and our
results  agree. The three body phase space calculation was done by computer.
\hfill\break\indent The matrix element squared may be written as a sum of
matrix element squared of left- and right-handed components due to the fact the
electrons are massless: $$|{\cal M}|^2 = |{\cal M}_L|^2 + |{\cal
M}_R|^2.\eqno(A.1)$$ This is further split into a non-interference term and an
interference term, the interference term coming from the cross traces of the
diagrams, and the diagrams with $p_1$ interchanged with $p_2$.  Thus  $$|{\cal
M}_L|^2 = {1\over 2}\big(\;|{\cal M}_L^{NI}|^2 + |{\cal M}_L^I|^2 +
(p_1\rightleftharpoons p_2)\;\big).\eqno(A.2)$$
The expression for the non-interference term is then given by:
$$\displaylines{|{\cal M}_L^{NI}|^2 = -16e^2[-eN'_{11} + {{c^e_L g
N'_{12}}\over{ cos\theta_W }} ]^4\hfill\cr
\quad\times\Bigg [\; |a_L|^2(p_1\cdot p_e)(p_2\cdot
k_e) -{{4(p_e\cdot p_\gamma )(p_1\cdot p_\gamma )(p_2\cdot k_e)
}\over{s^2|P_{K2L}|^2}} -{{4(k_e\cdot p_\gamma )(p_1\cdot p_e)(p_2\cdot
p_\gamma
) }\over{t^2|P_{P1L}|^2}}
\hfill\cr  \qquad
+2\Big(Re({a_L\over{P^*_{K2L}}})\cdot p_e\Big)
{{(p_1\cdot p_\gamma )(p_2\cdot k_e )}\over{s}}
-2\Big(p_1 \cdot Re({a_L\over{P^*_{K2L}}})\Big){{(p_e\cdot p_\gamma )(p_2\cdot
k_e )}\over{s}}
\hfill\cr \qquad
-2\Big(Re({a_L\over{P^*_{P1L}}})\cdot k_e\Big) {{(p_2\cdot p_\gamma
)(p_1\cdot p_e )}\over{t}}
+2\Big(p_2 \cdot Re({a_L\over{P^*_{P1L}}})\Big)
{{(k_e\cdot p_\gamma )(p_1\cdot
p_e )}\over{t}}\hfill(A.3)\cr
\qquad +\epsilon_{\alpha\beta\sigma\rho}\Big\{\quad
[p_1^\alpha p_e^\beta (k_e \cdot
p_2 ) + p_2^\alpha p_e^\beta (k_e \cdot p_1 )] p_{\gamma}^{\sigma}
Im({{a_L^{\rho}}\over{sP^*_{K2L}}}) \hfill\cr
\qquad\quad +[p_1^\alpha k_e^\beta (p_e \cdot p_2 ) + p_2^\alpha
k_e^\beta (p_e \cdot p_1 )] p_{\gamma}^{\sigma}
Im({{a_L^{\rho}}\over{tP^*_{P1L}}})\hfill\cr
\qquad\quad+p_1^{\alpha}p_2^{\beta}[
 -k_e \cdot p_{\gamma} p_e^{\sigma} Im({{a_L^{\rho}}\over{sP^*_{K2L}}})
+\Big(k_e \cdot Im({{a_L}\over{sP^*_{K2L}}})\Big) p_e^{\sigma}p_{\gamma}^{\rho}
+ p_e \cdot p_{\gamma} k_e^{\sigma} Im({{a_L^{\rho}}\over{tP^*_{P1L}}})
\hfill\cr\qquad\qquad\qquad\quad
-\Big(p_e \cdot Im({{a_L}\over{tP^*_{P1L}}})\Big)
k_e^{\sigma}p_{\gamma}^{\rho} + k_e \cdot p_e p_{\gamma}^{\sigma}
Im({{a_L^{\rho}}\over{sP^*_{K2L}}}-{{a_L^{\rho}}\over{tP^*_{P1L}}})]
\;\Big\}\;\Bigg ]. \hfill   }
$$
The interference term is given by:\vfill\break
$$\displaylines{|{\cal M}_L^{I}|^2 = +8e^2[-eN'_{11} + {{c^e_L g
N'_{12}}\over{ cos\theta_W }} ]^4 \tilde{m}^2\;\times\;\Bigg
[ \quad Re(a_L\cdot a'^*_L)(k_e\cdot p_e)
\hfill\cr\qquad\qquad
\eqalign{&-{{4(p_e\cdot p_\gamma )(k_e\cdot p_\gamma)}\over{s^2}}
Re\Big({1\over{P_{K2L}P_{K1L}^*}}\Big)
-{{4(p_e\cdot p_\gamma )(k_e\cdot p_\gamma)}\over{t^2}}
Re\Big({1\over{P_{P1L}P_{P2L}^*}}\Big)
\hfill\cr
&+2 Re\Big({{a_L}\over{sP^*_{K1L}}} + {{a_L}\over{tP^*_{P2L}}}\Big) \cdot p_e\>
(k_e\cdot p_\gamma)
- 2 Re\Big({{a_L}\over{sP^*_{K1L}}} + {{a_L}\over{tP^*_{P2L}}}\Big) \cdot k_e\>
(p_e\cdot p_\gamma)\hfill}\hfill(A.4)
 \cr
\qquad\qquad\qquad -2\epsilon_{\alpha\beta\sigma\rho}
\Big\{ \; p_e^\alpha k_e^\beta p_\gamma^\sigma\;
Im\Big( {{a_L^\rho}\over{sP^*_{K1L}}} - {{a_L^\rho}\over{tP^*_{P2L}}}\Big)\;
\Big\}\;\quad\Bigg ].\hfill
 }
$$
In the above formulas, $\tilde{m}$\ is the LSP mass, $m_{\tilde{e}_{L/R}}$\
the mass of the left/right-handed scalar electron,
  $$\displaylines{
\hfill s = (p_e+p_\gamma)^2,\hfill (A.5)\cr\hfill t =
(k_e-p_\gamma)^2,\hfill (A.6)\cr \hfill P_{K\,1/2\,L} = (k_e+p_{1/2})^2 -
{m}^2_{\tilde{e}_L} + i\Gamma_{\tilde{e}_L}{m}^2_{\tilde{e}_L},\hfill(A.7)\cr
\hfill P_{P\,1/2\,L} =
(p_e-p_{1/2})^2 - {m}^2_{\tilde{e}_L} +
i\Gamma_{\tilde{e}_L}{m}^2_{\tilde{e}_L},\hfill(A.8)\cr
\hfill a_L^\mu
= {{2p_e^\mu}\over{sP_{K2L}}}
+ {{2k_e^\mu}\over{tP_{P1L}}}
+ {{p_e^\mu + k_e^\mu - p_1^\mu + p_2^\mu}\over{P_{P1L}P_{K2L}}},\hfill(A.9)
\cr
\noalign{\hbox{and}}
\hfill a'^\mu_L = a^\mu_L (p_1\rightleftharpoons p_2)
= {{2p_e^\mu}\over{sP_{K1L}}}
+ {{2k_e^\mu}\over{tP_{P2L}}}
+ {{p_e^\mu + k_e^\mu + p_1^\mu - p_2^\mu}\over{P_{P2L}P_{K1L}}}.\hfill
(A.10)}$$ Similar expressions hold for the right-handed terms with
L$\rightarrow$R.  The Z coupling constants, $c^e_{L/R}$\ are given by
$$\eqalignno{\eqalign{c_L^e=&-{{1}\over{2}} + \sin ^2\theta_W,\cr
c_R^e=&\sin ^2\theta_W.}&&(A.10)}$$\eject
{\bf References}\vskip.2cm
\item{[\ 1]}M.C. Gonzalez-Garcia et al.,``Isosinglet Neutral
Heavy Lepton Production in High-Energy $e-\gamma$\ collisions'',
MAD/PH/686, IFT-P-044/91.
\item{[\ 2]}E. Boos et al., Phys. Lett. {\bf 273B} (1991), 173.
\item{[\ 3]}G. Couture et al., Phys. Rev. {\bf D39} (1989), 3239.
\item{[\ 4]}S.Y. Choi and F. Schrempp, Phys. Lett. {\bf 272B}.
 (1991), 149
\item{[\ 5]}E. Yehudai, Phys. Rev. {\bf D44} (1991), 3434.
\item{[\ 6]}L. Bento and A. Mour\~ao, Z. Phys. {\bf C37} (1988),587.
\item{[\ 7]}A. Goto and T. Kon,
Europhys. Lett. {\bf 13} (1990), 211.
\item{[\ 8]}I.F. Ginzburg et al., Nucl. Inst. Meth. {\bf 205}
(1983), 47, Nucl. Inst. Meth. {\bf 219} (1984), 5.
\item{[\ 9]}
R. Blankenbecler and S.D. Drell, Phys. Rev. Lett.
{\bf 61} (1988), 2324.
\item{[  10]}H.E. Haber and G.L. Kane, Phys. Rep.
{\bf 117} (1985), 75.
\item{[  11]}L. Roszkowski, Phys. Lett. {\bf 252B} (1990),471.
\item{[  12]}L.E. Iba\~nez and C. L\'opez, Nucl. Phys.
{\bf B233} (1984), 511.
\item{[  13]}A. Goto and T. Kon, ``Supersymmetric Particle
Production at TeV $e\gamma$\ collider'', ITP-SU-92/01.
\hfill\break\vfill
\eject

{\bf Figure Captions}\vskip.1cm
\item{Fig. 1\phantom{\ a)}}Feynman diagrams for producing
the LSP in electron-photon collisions.
\item{Fig. 2\phantom{\ a)}}Feynman rules for the couplings used
in Fig. 1
\item{Fig. 3\ a)} The differential cross section
$\displaystyle{{{d\sigma}\over{dE_e}}}$\
 for the CME
$\sqrt{s}=100$\ GeV: The solid line is the Standard Model result,
the dashed line is the MSSM result with $m_{\tilde e}=50$\ GeV,
the dotted line is $m_{\tilde e}=75$\ GeV and the dash-dotted line is
$m_{\tilde e}=100$\ GeV.  The dash-triple dotted line is the Standard
Model result for a right-handed polarized electron beam.  For the
supersymmetric processes there is no chiral dependence under our
assumptions. The mass of the LSP was taken to be 30 GeV.
\item{\phantom{Fig. 4}\ b)} Same as a) except for
 the differential cross section
$\displaystyle{{{d\sigma}\over{dcos\theta_e}}}$.
\item{\phantom{Fig. 4}\ c)} The differential cross section
$\displaystyle{{{d\sigma}\over{dE_e}}}$\
with the photon energy distribution folded in for an $e^+e^-$\ CME
$\sqrt{s}=100$\ GeV and a laser energy of 3.5 keV: Labeling the
same as in a).
\item{\phantom{Fig. 4}\ d)} Same as c) except for
 the differential cross section
$\displaystyle{{{d\sigma}\over{dcos\theta_e}}}$.
\item{Fig. 4\ a)} The differential cross section
$\displaystyle{{{d\sigma}\over{dE_e}}}$\
 for the CME
$\sqrt{s}=500$\ GeV:
The solid line is the SM result,
the dashed line is the MSSM result with $m_{\tilde e}=50$\ GeV,
the dotted line is $m_{\tilde e}=100$\ GeV and the long-dash-dotted line is
$m_{\tilde e}=250$\ GeV. The dash-triple-dotted line is the SM result for a
right-handed polarized electron beam and the
short-dash-dotted line is for $m_{\tilde e}=250$\ GeV with right-handed
polarization.  For smaller $m_{\tilde e}$\ there is no chiral dependence.
The mass of the LSP was taken to be 30 GeV.
\item{\phantom{Fig. 4}\ b)} Same as a) except for
 the differential cross section
$\displaystyle{{{d\sigma}\over{dcos\theta_e}}}$.
\item{\phantom{Fig. 6}\ c)} The differential cross section
$\displaystyle{{{d\sigma}\over{dE_e}}}$\
with the photon energy distribution folded in for an $e^+e^-$\ CME
$\sqrt{s}=500$\ GeV and a laser energy of 1.17 keV:
Labeling the same as in a).
The right-hand polarized case for $m_{\tilde e}=250$\ GeV,
 is very similar to the
unpolarized case and has been omitted for aesthetic reasons.
\item{\phantom{Fig. 4}\ d)} Same as c) except for
 the differential cross section
$\displaystyle{{{d\sigma}\over{dcos\theta_e}}}$.
\item{Fig. 5\ a)} The differential cross section
$\displaystyle{{{d\sigma}\over{dE_e}}}$\
 for the CME
$\sqrt{s}=1$\ TeV: Labeling the same as in Fig. 4 a).
\item{\phantom{Fig. 4}\ b)} Same as a) except for
 the differential cross section
$\displaystyle{{{d\sigma}\over{dcos\theta_e}}}$.  The theta dependence
for $m_{\tilde e}=50$\ GeV is much the same as for $m_{\tilde e}=100$\ GeV
and has been omitted for aesthetic  purposes.
\item{\phantom{Fig. 8}\ c)} The differential cross section
$\displaystyle{{{d\sigma}\over{dE_e}}}$\
with the photon energy distribution folded in for an $e^+e^-$\ CME
$\sqrt{s}=1$\ TeV and a laser energy of 1.17 keV:
Labeling the same as in a).
\item{\phantom{Fig. 4}\ d)} Same as c) except for
 the differential cross section
$\displaystyle{{{d\sigma}\over{dcos\theta_e}}}$.  Again the case of
$m_{\tilde e}=50$\ GeV  has been omitted for aesthetic  purposes.
\end